\documentstyle[12pt]{article}

\catcode`\@=11
\long\def\@makefntext#1{
\protect\noindent \hbox to 3.2pt {\hskip-.9pt
$^{{\ninerm\@thefnmark}}$\hfil}#1\hfill}		

\def\@makefnmark{\hbox to 0pt{$^{\@thefnmark}$\hss}}  

\def\ps@myheadings{\let\@mkboth\@gobbletwo
\def\@oddhead{\hbox{}
\rightmark\hfil\ninerm\thepage}
\def\@oddfoot{}\def\@evenhead{\ninerm\thepage\hfil
\leftmark\hbox{}}\def\@evenfoot{}
\def\sectionmark##1{}\def\subsectionmark##1{}}

\newcounter{sectionc}\newcounter{subsectionc}\newcounter{subsubsectionc}
\renewcommand{\section}[1] {\vspace{0.6cm}\addtocounter{sectionc}{1}
\setcounter{subsectionc}{0}\setcounter{subsubsectionc}{0}\noindent
	{\bf\thesectionc. #1}\par\vspace{0.4cm}}
\renewcommand{\subsection}[1] {\vspace{0.6cm}\addtocounter{subsectionc}{1}
	\setcounter{subsubsectionc}{0}\noindent
	{\it\thesectionc.\thesubsectionc. #1}\par\vspace{0.4cm}}
\renewcommand{\subsubsection}[1]
{\vspace{0.6cm}\addtocounter{subsubsectionc}{1}
	\noindent {\rm\thesectionc.\thesubsectionc.\thesubsubsectionc.
	#1}\par\vspace{0.4cm}}

\newcounter{appendixc}
\newcounter{subappendixc}[appendixc]
\newcounter{subsubappendixc}[subappendixc]

\renewcommand{\appendix}[1] {\vspace{0.6cm}
        \refstepcounter{appendixc}
        \setcounter{figure}{0}
        \setg{table}{0}
        \setcounter{equation}{0}
        \renewcommand{\thefigure}{\Alph{appendixc}.\arabic{figure}}
        \renewcommand{\thetable}{\Alph{appendixc}.\arabic{table}}
        \renewcommand{\theappendixc}{\Alph{appendixc}}
        \renewcommand{\theequation}{\Alph{appendixc}.\arabic{equation}}
        \noindent{\bf Appendix \theappendixc #1}\par\vspace{0.4cm}}

\newcounter{itemlistc}
\newcounter{romanlistc}
\newcounter{alphlistc}
\newcounter{arabiclistc}

\newcommand{\fcaption}[1]{
        \refstepcounter{figure}
        \setbox\@tempboxa = \hbox{\tenrm Fig.~\thefigure. #1}
        \ifdim \wd\@tempboxa > 6in
           {\begin{center}
        \parbox{6in}{\tenrm\baselineskip=12pt Fig.~\thefigure. #1}
            \end{center}}
        \else
             {\begin{center}
             {\tenrm Fig.~\thefigure. #1}
              \end{center}}
        \fi}

 1
 1
 1

\font\tenrm=cmr10

\font\ninerm=cmr9

\begin{document}
\raggedbottom
\def \TS {$^3S_1$}
\def \SS {$^1S_0$}
\def \TPone {$^3P_1$}
\def \TP0 {$^3P_0$}
\def \TPtwo {$^3P_2$}
\def \SPone {$^1P_1$}
\def \Pii {$\Pi_{\alpha\beta}(s)$}
\def\ss {$s\bar s$}
\def\an {$a_0(980)$}
\def\fs {$f_0(980)$}
\def\fn {$f_0(1300)$}
\def\Kst {$K^*_0(1430)$}
\def \nn {$n\bar n$}
\def\qq {$q\bar q$}
\def\non {non-\qq}
\def\ud {$u\bar d$}
\def\su {$s\bar u$}
\def \K {$K^*$}
\def\uu { $(u\bar u+d\bar d)/\sqrt 2$}
\def\0++ {$0^{++}$}
\def\KK {$K\bar K$}

\hfill HU-SEFT R 1995-05 (for ZPC)

\bigskip

\centerline{\bf UNDERSTANDING THE SCALAR  MESON $q\bar q$ NONET}
\bigskip
\centerline{ NILS A. T\"ORNQVIST}
\bigskip
{\it \centerline{University of Helsinki, Research Institute for
High Energy Physics}
\centerline{PB 9, Siltavuorenpenger 20, Fin-00014 Helsinki, Finland}}
\bigskip
\begin{quote}
\small
\centerline{ABSTRACT}
It is shown that one can fit the available data on the $a_0(980)$, $f_0(980)$,
$f_0(1300)$ and $K^*_0(1430)$ mesons as a distorted $0^{++}$ \qq\ nonet using
very few (5-6) parameters and an improved
version of the unitarized quark model. This
includes all light two-pseudoscalar  thresholds, constraints from Adler
zeroes, flavour symmetric couplings,
unitarity and physically acceptable analyticity.
The parameters include  a bare $u\bar u$ or $d\bar d$ mass,
an over-all coupling constant, a cutoff and a strange quark mass of $ 100$ MeV,
which is in accord with expectations from the quark model.

 It is found that in particular for the \an\
and \fs\ the \KK\ component in the wave function is large, i.e., for a large
fraction  of the time the \qq\ state is transformed into a  virtual \KK\
pair.  This \KK\  component, together with a similar component of
$\eta' \pi$ for the \an , and  $\eta\eta,\eta\eta'$ and
$\eta'\eta'$  components for the \fs , causes
the substantial shift to a lower mass than what is
naively expected from the \qq\ component alone.

Mass, width and  mixing parameters, including sheet and pole positions,
of the four resonances are given, with a detailed pedagogigal discussion
of their meaning.
\end{quote}

\vskip 0.6cm
{\bf 1 Introduction.}
\vskip 0.5cm
As has often been stated in many reviews\cite{rev}
 our present understanding of the light
meson mass spectrum is in a deplorable state, especially when one considers the
vast amount of data that has been available already for quite some time,
and that  QCD {\it in principle} should solve the hadron spectrum. This is
mainly because of the fact that the expectations of most "QCD inspired
quark models" fail so dramatically for the scalar mesons, - the Higgs
bosons of the hadronic sector.

For the other \qq\ nonets such as the \TS\ $(\rho,\omega, K^*,\phi ),$
\TPtwo\ $(a_2, f_2, K^*_2, f'_2)$, \TPone\ $(a_1, f_1,$$K_{1A}, f'_1)$,
\SPone\ $(b_1,$$h_1, K_{1B}, h'_1)$ and even the $^1S_0$ $(\pi,\eta ,K,\eta')$
the naive quark model works reasonably
well as a rough first approximation. Therefore, few authors doubt that
they should be classified as
\qq\ states. Certainly, also here  there are some
"second order effects" such as the observed deviations from
ideally mixed states (i.e.,  mass splittings like
$\rho - \omega$ or $\Delta - N$ and
mixing angles like $\phi -\omega$ or $K_{1A} - K_{1B}$),
for which one has not yet reached
consensus as to their full origin. Many authors believe these
deviations are mainly due to gluonic intermediate states, while
others\cite{lipkin}, including  myself,  believe the dominant
effects come from hadronic
loops like $\phi\to K\bar K\to \omega$ etc.

But, for the lightest scalars the \an , \fs , \fn , and \Kst\ one has not even
reached a clear consensus as to their true nature.
Are some of these \qq\ or \KK\ bound states?  Or
is one of them possibily a glueball? Many authors today
believe the \an\
and the \fs\ to be \KK\ bound states\cite{rev,KK}.
 This seems, at first,  to be a natural assumption,
since they lie just a little below the \KK\ threshold. If so, the I=1 and the
\ss\ state must be sought for at higher masses. And indeed, there are now
candidates for such states: the $a_0(1450)$ of Crystal barrel\cite{Xala0}
 (or possibly the questionable $a_0(1320)$ of GAMS\cite{GAMS})
 for the I=1 state, and the LASS $f_0(1525)$ for the \ss\ state. However,
flavour symmetric couplings
(which works rather well for the  established nonets)
would require\cite{goar} that their widths
should be at least 500MeV, which is much larger than the observed widths of
these candidates. Furthermore,
these can have other interpretations (as radial excitations, meson-meson bound
states, glueball or threshold effects) like many of the other
observed scalars in
the much too overpopulated 1370-1720MeV region, where many
I=0 candidates\cite{PDG},
$f_0(1370)$, $f_0(1450)\cite{WAf0},\ f_0(1525),\ f_0(1590),
\ f_{0/2}(1710)$, do not  find a place in the \qq\ quark model.
One of these could be a glueball,  another
a deuteronlike state (deuson)\cite{deuson}
etc. In order to reach a better conclusion of their true nature it would of
course be very helpful if at least the lightest scalar \qq\ nonet would be
resolved.

It  is fundamental also in many other
respects to have a good model of the scalars,
in particular for understanding chiral symmetry breaking,
nuclear forces and of course confinement.
For chiral symmetry breaking it would be important to know:  Where is the
sigma meson? Is it the \fs , or the \fn, or must its mass  be
pushed to infinity? How does
the Nambu--Jona-Lasinio mechanism\cite{NJL}
work for the light spectrum? Is the
pion both the Nambu-Goldstone boson and the I=1 $^1S_0$
 \qq\ meson, as most authors believe, or does one have
to look for the \qq\ pion at higher masses, as questioned e.g.,
by Georgi and Manohar\cite{manohar}.

For nuclear forces, and for the understanding of the $\Delta I=1/2$ enhancement
in $K\to 2\pi$ and $K_{l4}$ decay\cite{meissner2},
 one would like to have a very light
$\sigma $, in the range of 600-900 MeV, coupling strongly to $\pi\pi$; a meson
which does not seem to
exist.  The lightest scalar is the \fs , which behaves more like an \ss\
or a \KK\ state coupling weakly to $\pi\pi$, while the \fn\ seems  too
heavy. In this paper  I shall suggest a new solution to this old question.

Finally, for confinement the scalars play an important role in building
through tadpole diagrams the hadronic bag, within which the quarks reside.
They are important for  the confinement energy, for the \qq\ condensate,
and for  the difference between constituent and chiral quark masses. Thus,
they  are  crucial for the understanding of all hadronic masses.

Many authors
 have recently studied some of the scalars we discuss here, but
generally these have tried to fit at the same time only one or two
of them and not the
whole flavour nonet simultaneously. Bugg et al.\cite{bugg} and Morgan and
Pennington\cite{morgan} have made detailed fits with many parameters
 to $\pi\pi$ amplitudes
using general techniques of unitarity, analyticity and the K-matrix.
 Janssen et
al.\cite{speth} studied the \an\ and \fs\ in a meson exchange model for
$\pi\pi$ and $\pi\eta$ interactions, Achasov\cite{NN} studied the low energy
$\pi\pi$ data also with meson exchange and Adler zero constraints.
Kaminsky et al. \cite{lesniak} studied the \an\ and \fs\ using a coupled
channel model concluding that the \fs\ is a "\KK\ molecule". Earlier many
authors\cite{menessier} have discussed unitarity and analyticity for
the hadron spectrum.

The results presented here is a very much improved calculation and discussion
of a short
letter\cite{UQMscal} 14 years ago. In particular, I have included
constraints from Adler zeroes, which considerably improve the agreement with
data close to thresholds without increasing the number of parameters.

In the following I first discuss the general ingredients and properties
of  the unitarized
quark model (UQM) in Sec.2. Although this section does contain new
results and new material to resonance phenomena and the UQM, it can perhaps
be skimmed by those who just want to understand the fits. It is written
in a rather pedagogical way
and emphasizes some facts, which are often forgotten by many model builders.
Sec.3 is the central chapter of this paper, where  the actual
application and fits to
the scalar nonet are presented while in the concluding remarks, Sec.4,
some comments on  these results are discussed.

\vskip 0.5cm
{\bf 2. The unitarized quark model.}
\vskip 0.4cm

The UQM incorporates unitarity and physically acceptable
analyticity with resonances in a
way, by which one maintains a simple and transparent physical
interpretation of the introduced resonance  parameters.
It is a kind of advanced form of the
classical work of Weisskopf and Wigner\cite{weissk}.
The UQM was applied to many different hadrons\cite{UQM}, and it can explain
 the signs\cite{goar} and magnitudes of deviations from ideal mixing, and many
mass splittings. Particularily significant is the large splitting between
 $\Upsilon (5S)$ and $
\Upsilon (4S)$, which cannot be understood in single channel potential
models (where the predicted splitting is over 50 MeV
 too small) nor by gluonic exchange. At present there is no other mechanism
than hadronic shifts from the loops $\Upsilon(nS)
 \to  D \bar D,$ $D \bar D^*$ etc.
$\to \Upsilon (mS)$\cite{NATup}, which can account for this large splitting.
\vskip 0.4cm
{\bf 2.1 General formulation}
\vskip 0.2cm
  In the UQM one writes for the partial wave amplitudes (PWA) in "Argand units"
a factorized matrix form:

\begin{equation}
  A_{ij}(s) =
T_{ij}(s)(k_ik_j)^{\frac 1 2 } = \sum_{\alpha\beta}
G^\dagger(s)_{i\alpha}
P_{\alpha\beta}(s) G_{\beta j}(s) \  , \label{aij}
\end {equation}

\noindent
where one sums over the resonance indices $\alpha$ and $\beta$
(for e.g., \fs , \fn\ etc.)  and
where $i$ and $j$ denote the two-body thresholds (e.g., $\pi\pi, K\bar K$
etc.).
 The matrices $G_{\alpha i}(s)$
include coupling constants ($g_{\alpha i}$), phase space (and angular momentum)
factors, and form factors. We write
for real $s$:

\begin{equation}
G^2_{\alpha i}(s)=g^2_{\alpha i}\frac {k_i(s)}{\sqrt s}F^2_{\alpha
i}(s)\theta (s-s_{th,i})
\ , \ \ s_{th,i}= (m_{A_i}+m_{B_i})^2 \ , \label{gs}
\end{equation}

\noindent
where $k_i(s)=[\lambda(s,m^2_{A_i},m^2_{B_i})/s]^{1/2}/2$ is the cm momentum
of the two intermediate particles $A_i$ and $B_i$. The form factors
 $F_{\alpha i}(s)$ are at this stage still
quite arbitrary, except for the fact that we shall require them to vanish
sufficiently fast at $\infty$ (In the following we generally suppress the index
$\alpha$ on the $F_{\alpha i}$, since in our application in this paper we
assume this does not depend on the resonance). One expects them
to be smooth functions of $s$,
which include  angular momentum barriers, radial nodes, and in principle
the left hand cuts. In the quark pair creation
model\cite{olivier} they are given by an overlap of the three hadronic
wave functions
multiplied by a matrix element for the \qq\ pair creation. In fact, the
$F_i(s)$
include most of the model depencence of our scheme.

For \qq\ resonances the propagator matrix $P_{\alpha\beta}(s)$ depends on
 the bare mass parameters $m_{0,\alpha}$, and on the mass shifts
 $Re\Pi_{\alpha\beta }(s)$,
and the width-like functions $Im\Pi_{\alpha\beta}(s)$,
which together determine the analytic
vacuum polarization functions $\Pi_{\alpha\beta } (s)$.

\begin{equation}
 P^{-1}_{\alpha\beta}(s)= (m_{0,\alpha}^2 -s)\delta_{\alpha\beta}
+\Pi_{\alpha\beta}(s)  \ , \label{ps}
\end{equation}
\noindent

The unitarity condition,  $(A-A^\dagger)/(2i) = AA^\dagger ,$
takes a  very simple
form determining  the imaginary part of $\Pi_{\alpha\beta}(s)$:

\begin{equation}
Im \Pi_{\alpha\beta}(s) = - \sum_i G_{\alpha i}(s)G_{i \beta }^\dagger (s)=
-\sum_ig_{\alpha i}g_{\beta i} \frac{k_i(s)}{\sqrt s}
F_i^2(s) \theta (s-s_{th,i}) \ .
\label{impi}
\end{equation}
\noindent
Since the functions $\Pi_{\alpha\beta}(s)$ are  analytic functions
 with only right hand cuts,
we can write  dispersion relations for the real parts $Re\Pi_{\alpha\beta}(s)$:

\begin{equation}
Re\Pi_{\alpha\beta}(s) =\frac {1}{\pi} {\cal P} \int_{s_{th,1}}^{\infty}
\frac {Im\Pi_{\alpha\beta}(s')} {(s'-s)} ds' \ . \label{repi}
\end{equation}

\noindent These need no subtractions,
since we require that the hadronic form factors $F_i(s)$  make
$Im\Pi_{\alpha\beta}(s)$ go to zero sufficiently fast  at
infinity because hadrons have finite size.  Thus the integrals are finite and
we need not add any
polynomial to $Re\Pi_{\alpha\beta}(s)$
apart from the $m_{0,\alpha}^2 -s$ term, which we already
included in $P^{-1}$ in order to have the \qq\ resonances as
CDD\cite{CDD} poles. By defining $Re\Pi_{\alpha\beta}(s)$ through the
dispersion relation one automatically satisfies  physically correct
analytic properties, i.e., one gets no spurious poles nor cuts and right
asymptotic behaviour. (See discussion in Sec.~2.7. below).
Thus, once we have a model for the $G_{\alpha i}(s)$ and the bare masses,
the PWA can be calculated.

By having the functions $\Pi_{\alpha\beta}(s)$ in the inverse propagator one
automatically sums over all iterated loop diagrams of the Born terms
(see the diagrams in Fig.~1a,b), such that the amplitude includes an infinite
set of diagrams
of the form shown in Fig.~1c. In Fig.~1 we have included, in addition to
the \qq\ resonance terms also contact terms (Fig.~1b) to be discussed in
subsection 2.5.

\begin{figure}
\vspace  {9.5cm}
\fcaption{ The Born term for (a) the bare \qq\ resonances, for (b) the
contact terms, and (c) the loops summed by the functions $\Pi_{\alpha\beta}(s)$
in the inverse propagator.
}
\end{figure}
\vskip 0.5cm

{\bf 2.2 A single resonance}
\vskip 0.4cm
\begin{figure}
\vspace{19.5cm}
\fcaption{ (a) The real and the imaginary parts of the function
$m^2(s)=m^2_{0i}+\Pi (s)$  for the
$K^*_0(1430)$ resonance plotted as a function of $\sqrt s$.
Note the strong cusp at the $K\eta '$ threshold and that
the $K\eta$ threshold essentially decouples because of the small
coupling constant.
The dashed curve is $s$, which crosses the running mass
Re$[m^2(s)]$ at the BW mass.

(b) As in (a) but for the
\an\ resonance. Note that the three thresholds have similar coupling
strengths and that they lie much closer together than
in (a). Therefore the large mass shift for \an . The dashed curve is $s$,
which crosses Re$[m^2 (s)]$ at the BW mass.
 }
\end{figure}
\vskip 0.5 cm

Some special cases are instructive. For a single resonance $P(s)$ is
one-dimensional, and assuming real $G_{i}(s)$
(as can quite generally be done for $2 \to 2$ particle amplitudes) the PWA
takes the form:

\begin{equation}
A_{ij} = \frac {G_i(s)G_j(s)}{m_0^2+Re\Pi (s) -s- i\sum_iG^2_i(s)} =
\frac {m_{BW}(\Gamma_i(s)\Gamma_j(s))^{\frac{1}{2}}}
{m^2(s)-s-im_{BW}\Gamma_{tot}(s) }\ .\label{BW}
\end{equation}
\noindent
 One recovers a generalization of the familiar Breit-Wigner
form in the second expression of eq.(\ref{BW}), where
one has put $\Gamma_i(s)=G^2_i(s)/m_{BW}$. Here $m^2(s)=m_0^2+Re\Pi (s)$
is the "running squared  mass", which is given by the bare squared mass plus
the
generally negative
 mass shift $Re\Pi (s)$. This function is approximately constant
only in special situations, e.g.,
if one is far from all thresholds. For S-waves the $s$-dependence
is particularily important, since $Re \Pi (s)$ has square root cusps at each
threshold, and near such a threshold there is a dramatic $s$-dependence.
See Fig.~2a,b where we display the running mass and
$Im\Pi (s)$ for the  $K^*_0$ and for the $a_0$ as obtained
in the fit presented  in the next section.
Note in particular the strong  cusp at the $K\eta'$ threshold. Since the
$K^*_0(1430)$ lies sufficiently below this threshold the $s$-dependence
of $m^2(s)$ is here not so crucial as it is
for the \an\ and for the \fs\
resonances, which lie essentially right at the \KK\ threshold. One sees
from Fig.~2b that for the \an\ all three thresholds
 $\pi\eta$, \KK , and $\pi\eta'$
lie close to each other. They all contribute to a large $a_0(980)$ mass
shift and a strong $s$-dependence in the running mass.

\vskip 0.5cm

{\bf 2.3 Two and more resonances}
\vskip 0.4cm

 When one considers two resonances the propagator matrix becomes
two-dimen\-sio\-nal
and has important off-diagonal elements given by \Pii . In general, one
gets an $s$-dependent complex mixing angle between the two states, since the
mass and propagator matrices are diagonalized by a complex orthogonal matrix.
This generates e.g.,
OZI violation although the bare states are assumed ideally mixed \qq\ states.

A special, instructive case is obtained if one has only one threshold,
or if the coupling constants of the two resonances are proportional
to each other, $G_{2i}(s) =\alpha(s) G_{1i}(s)$.
Such a proportionality can hold when
the resonances have the same flavour content, say two $K^*_0$'s with flavour
symmetric couplings.
 Then  both bare states have couplings   proportional to the same $SU_{3f}$
Clebsch-Gordan coefficients, which are given by the general trace formulae:

\begin{equation}
g_{\alpha,i} = g Tr [{\cal M}_{A_i}{\cal M}_{B_i}{\cal M}_{C_{\alpha}}
]_{C_{A_i}C_{B_i}C_{C_{\alpha}}}\ , \label{flavour}
\end{equation}
\noindent where $A_i,\ B_i,\ ,C_{\alpha}$ stand for the three mesons involved
 at the vertex $C_\alpha\to A_i+B_i$, and
${\cal M}_{A_i}$ is the 3x3 flavour matrix for meson $A_i$.
Either the symmetric or the
antisymmetric trace is to be taken, depending on the sign of the product
the three charge conjugation quantum numbers. For the
$0^{++}\to 0^{-+}\ 0^{-+}$  vertices, which we discuss in this paper it
is thus the symmetric trace.
Then after some  algebra  one can reduce\footnote
{The result looks almost like magic
if one starts from eq. (\ref{aij}), but it
is  easily found starting with the $ K$ or even better with the $B$
matrix defined in Sec. 2.7.}
the expression (\ref{aij}) to:

\begin{equation}
A_{ij}=\frac{[1+\alpha^2(s)]G_{1i}(s)G_{1j} (s)}{(m^2_{0,1}-s)(m^2_{0,2}-s)/
(\bar m^2(s)-s) +\Pi (s)} \ ,\label{twoBW}
\end{equation}

\noindent where
\begin{eqnarray}
\alpha(s)&=&G_{2i}(s)/G_{1i}(s)= g_{2i}F_{2i}(s)/[g_{1i}F_{1i}(s)] \ ,
\label{alpha} \\
\bar m^2(s) &=& \frac { m^2_{0,1}\alpha^2(s)+m^2_{0,2}}  {1+\alpha^2 (s)}
 \label{twores} \ ,\\
Im\Pi (s) &=& -[1+\alpha^2(s)]\sum_i G_{1i}^2(s)\ , \\
Re\Pi (s) &=& \frac {1} {\pi} {\cal P}\int_{s_{th}}^\infty \frac
{Im\Pi (s')ds'}{s'-s} \ .\\
\end{eqnarray}

This shows explicitely how
resonances must be "added multiplicatively" because of
unitarity. Wheras the single resonance form (\ref{BW}) gives one loop in the
Argand diagram the two resonance form (\ref{twoBW}) gives two loops with
a zero in the amplitude at $s=\bar m^2(s)$. Note that the zero is unshifted,
which could be phrased as a theorem:
{\it A zero in the PWA in the physical
region remains a zero after unitarization. }

 I find it rather surprising that this quite
simple generalization (\ref{twoBW}) of the BW
formula to the case of two resonances is not found
in the literature, although it could be very useful
phenomenologically e.g., when studying two resonances
 in channels with nonzero flavour like $K\pi$, where
eq. (\ref{alpha}) can hold.

A special case of eq.(\ref{twoBW}) is also instructive, and this bears
some resemblance to the "S$^*$ effect",
i.e. the \fs\ discussed below: Imagine two
nearly degenerate resonances both with large couplings to common channels.
Unitarity will shift both masses and the phase shift will pass 90$^\circ$
for the first and 270$^{\circ}$ for the second resonance, close to each
other in energy. Thus, one gets  a narrow resonance structure
in the form of a dip between two broad bumps. How can this this come about?
In fact, when looking at the original form, eq.(\ref{aij}),
the $2\times 2$ mass matrix in the propagator, when diagonalized
[cf. next subsection 2.4,  eqs.(\ref{mdiag}-\ref{diagsum})]
results in a mixing between the two resonances, such that one of them nearly
decouples from the thresholds, while the other gets very large couplings
(cf. eq.(\ref{gprime}) below). Thus in terms of eigenvalues one has one
very broad and one very narrow resonance,
the latter producing a strong dip in the broad resonance bump.
Or in other words,
 two inherently broad resonances can together produce one very narrow one!

For  the \fs\ this mechanism is of course not the whole story.
Here the couplings are not proportional to each other,
 i.e., they do not satisfy eq.(\ref{alpha}), and the \KK\ threshold
plays a crucial role in bringing the two resonances close
to each other, but loosely speaking a variant of this mechanism is
operative. A somewhat similar phenomenon appears also for the two axial
$K_1$ mesons,
which are near $45^\circ$ mixtures of the $K_{1A}$ and $K_{1B}$,
 (which belong to the $1^{++}$ and
$1^{+-}$ nonets) such that one of them
nearly decouples from $K^*\pi$ and the other
decouples from $K\rho$ (cf. Katz and Lipkin\cite{lipkin}). Also in the late
sixties there were discussions of similar effects in the connection with
split resonances\cite{dothan}
(in particular the now well forgotten "split $a_2$").

For the more general case with two (or more) resonances and
many thresholds with different kinds of couplings, like  \uu\ and  \ss\
states, one does
not gain much insight  by  trying to reduce  the matrix form of the propagator
of eq. (\ref{aij}) algebraically. The "reduced" formula becomes
quite complicated because of the energy dependent complex
mixing induced.  But, fortunately eq. (\ref{aij}) is  as it stands already
quite transparent physically, and easy to compute numerically.

\vskip 0.5cm
{\bf 2.4 Breit-Wigner masses and widths, pole positions and resonance mixings.}
\vskip 0.4cm

For the single resonance case eq.(\ref{BW}) the Breit-Wigner
($90^\circ$) mass is given by the
$s$-value where $m_0^2+Re\Pi (m_{s})-s$ vanishes or by
\begin{equation}
m_{BW}^2 = m_0^2+Re\Pi (m_{BW}^2) \ ,
\label{mBW}
\end{equation}
\noindent while the widths could be defined
as $\Gamma_i=G^2_i(m^2_{BW})/m_{BW}$ as in eq.(\ref{BW}). However,  if the
slope of $Re\Pi (s)$ at the resonance
is large, one should  correct for this and absorb the slope into the
$s$ term by dividing both numerator and denominator in eq.(\ref{BW})
by the same term, and
define the  BW  widths  by

\begin{equation}
\Gamma_i^{BW} = \frac{ G^2_i(m^2_{BW})/m_{BW}  }
{  1-\frac{d}{ds} Re\Pi (m^2_{BW} )  } \ ,\hskip 2cm
\Gamma_{tot}^{BW} = \sum_i \Gamma_i^{BW} \  . \label{gamBW}
\end{equation}

This renormalization of the widths and couplings
(which is also familiar in field theory) has a clear physical interpretation:
Below each threshold the \qq\ pair produces virtual pairs of the two mesons,
and the probability for such pairs is proportional to $-\frac d{ds}\Pi_i(s)$
$=-\frac 1 \pi \int ds'Im\Pi (s') / (s-s')^2$.
The \qq\ wave function  obtains  meson-meson components $|A_iB_i>$:
\begin{equation}
|\psi > = {\big [}|q\bar q > + \sum_i  [-\frac d{ds} Re\Pi_i(s)]^{\frac 12}
 |A_iB_i>{\big ]}/[1+\sum_i  (-\frac d{ds} Re\Pi_i(s)]\ .
\end{equation}

Each $|A_iB_i>$ component has in configuration space
an exponential radial tail,
whose slope is inversely proportional to how much below the threshold the
state is. Thus for the \an\ and \fs\ the size of the \KK\
component grows both in spatial size and in absolute
 magnitude the closer to the threshold from below the resonance is.
The reduction of the width and coupling in eq. (\ref{gamBW}) is physically
due to the
fact that only the \qq\ component annihilates directly to $\pi\eta$
respectively
$\pi\pi$. The $|K\bar K>$ part is rather inert,
since it must first transform into \qq\ near the origin
 and then into $\pi\eta$ or $\pi\pi$.
Above the threshold the situation is very different; the \KK\ component
vanishes since it can simply  fall apart, and gives an absorptive part
to the wave function.
The slope of $Re \Pi (s)$ is here generally positive
implying that the BW width is in fact enhanced.

In the case when one has many resonances one can define the generalization
of BW masses and widths by first diagonalizing the mass matrix and propagator
by a complex orthogonal matrix ${\cal O}(s)$, which also rotates the
couplings which now become complex:

\begin{eqnarray}
m^2_{diag, \alpha} (s) &=& \sum_{\alpha '\beta '}
{\cal O}_{\alpha\alpha'}^{-1}(s) [ m_{0,\alpha '}^2\delta_{\alpha '\beta '}+
\Pi_{\alpha '\beta '} (s) ] {\cal O}_{\beta '\beta} (s)\ , \label{mdiag}\\
g'_{\alpha i}(s) &=& \sum_\beta {\cal O}_{\alpha \beta}^{-1}(s)g_{ \beta i }\ ,
 \label{gprime}\\
A_{ij} &=& \sum_\alpha \frac {g'_{\alpha i}(s)g'_{\alpha j}(s)}
{m^2_{diag,\alpha}(s) -s}F^2_i(s)\frac {k_i(s)}{\sqrt s} \label{diagsum}\ .
\end{eqnarray}

One can then define the  BW masses in the multiresonance case as the
energies where  $Re [m^2_{diag,\alpha} (s)]-s$ vanishes. This definition has
the
advantage that this mass in principle is the same in all channels $i$. We shall
here use this definition.
Other definitions, such as the energies where the phase of each term in
the sum (\ref{diagsum}) is 90$^\circ$, or the energies
where their absolute value
of each term is maximal, do not have this property. But, the analogy with the
single resonance case is not simple,
 because the coupling constants $g'_{ \alpha i}(s)$  are now
complex and energy dependent and furthermore, there is a background from other
resonance tails.
 As an example the mixing angle between the \ss\
and \uu\ components  in  the \fs\ or \fn\ is complex
and is strongly $s$-dependent.
It has a different value at the \fs\ than at \fn , and furtermore it will be
different  when evaluated at the BW mass, than at the pole position for the
same data and model.

The pole positions have the advantage that only here does the process
factorize into production and decay independently of the background.
These are determined by the complex $s$ value
and sheet number,
where the the whole inverse propagator $m_0^2+\Pi (s)-s$ vanishes,
or more  generally
where $\det [P^{-1}(s)]$ vanishes. For each threshold the number of sheets is
doubled, and the sheet number is determined  by the signs of $Im(k_i)$.
The same resonance can, in general, have several image poles on different
sheets, which considerably complicates matters
since the same resonance can have more than one nearby pole (See Morgan and
Pennington\cite{morgan} and \cite{natpole} for
a discussion). In our model for the \fs\ and \fn\ we have 5 thresholds,
which means there are 32 sheets, some  of which could even have 2 poles each!
 To find the nearest poles one must analytically continue \Pii\, defined above
at the first sheet only just above the real axis.  At the first sheet
we can calculate \Pii\ by generalizing eq.(\ref{repi})
to complex $s$ values on the first sheet ($I$)
by the Cauchy integral around the cut on the real $x$ axis:

\begin{equation}
\Pi_{\alpha\beta}^{I}(s) + \frac {1} {\pi} \int_{s_{th}}^\infty
\frac {Im\Pi_{\alpha\beta}(x)dx}{x-s} \ . \\
\end{equation}

\noindent This is discontinuous across the cut. To get to the
second sheet ($II$) one must add a term:

\begin{equation}
\Pi_{\alpha\beta}^{II}(s) = \Pi_{\alpha\beta}^{I}(s)+
2i\bar G_{\alpha 1}(s)\bar G_{\beta 1}(s) \ , \\
\end{equation}

\noindent
where $\bar G_{1\alpha }(s)\bar G_{1\beta }(s)$ is given by essentially
the same expression as eq. (\ref{gs}) or eq.(\ref{impi}),  but
now without the theta function, and defined also for  complex values of $s$.
Near the real axis $\Pi^{II}_{\alpha\beta}(s)$ has the opposite
sign compared to
$\Pi^{I}_{\alpha\beta}(s)$ in  the imaginary part above threshold.
Again below threshold, the cusp in the real part changes sign.
Explicitly the additional term is given by:

\begin{equation}
 i\bar G_{\alpha i}(s)\bar G_{\beta i}(s) =
-g_{\alpha i}g_{\beta i}
\sqrt{ -\lambda (s,m_{A_i}^2,m_{B_i}^2) }s^{-1} F_i^2(s) \ ,
\label{gbar}
\end{equation}

\noindent or with the Adler zero constraints and flavour symmetric couplings
(See sec. 2.5):

\begin{eqnarray}
i \bar G_{\alpha i }(s)\bar G_{\beta i}(s) =& -\gamma_{\alpha i}
\gamma_{\beta i}(s-s_{A,i}) \sqrt{ -\lambda (s,m_{A_i}^2,m_{B_i}^2) }&s^{-1}
F_i^2(s) \ , \ \ \ {\rm where}\\ \gamma_{\alpha i}=& \gamma
{\big [}Tr[{\cal M}_{A_i}{\cal M}_{B_i}{\cal M}_{C_\alpha}]_+{\big ]}
 \ \ . \ \ \ \ \ \null &\label{flavour2}
\end{eqnarray}

\noindent In order to get to
the third sheet ($III$) two terms from the first and second threshold
must be added

\begin{equation}
\Pi_{\alpha\beta}^{III}(s) = \Pi_{\alpha\beta}^{I}(s)+
2i\bar G_{ \alpha 1}(s)\bar G_{\beta 1}(s)+
2i\bar G_{\alpha 2}(s)\bar G_{\beta 2}(s) \ ,\\
\end{equation}
\noindent while to get to the fourth sheet one should only add the second term.
In general, with the signs of $Imk_i$ given by the sheet number
($n$) one should
add $2i\bar G_{\alpha j}(s)\bar G_{ \beta j}(s)$ for each threshold
 $j$ with negative $Im k_j$:
\begin{equation}
\Pi_{\alpha\beta}^{n}(s) = \Pi_{\alpha\beta}^{I}(s)+
2i\sum_j \bar G_{\alpha j}(s)\bar G_{\beta j }(s)\theta(-Im k_j) \ .\\
\end{equation}
\vskip 0.6cm

{\bf 2.5 Background and $t$ and $u$ channel exchange terms.}
\vskip 0.4cm

For the case of a  nonresonant background term as given e.g., by a
contact term (Fig.~1b) one still has an expression very similar
to the one above, but with the $m^2-s$ terms in the propagator
replaced by a constant, which could be chosen = 1. But,
in order to have the same  dimensions for the coupling constants
of the background as for the resonances we
put this constant =$\pm M^2$ (with  dimension GeV$^2$),
and with the sign allowing for
constructive or destructive interference. Then  for the contact
terms we have:

\begin{equation}
 P^{-1}_{\alpha\beta,{\rm contact}}(s)=
\pm M^2_\alpha\delta_{\alpha\beta}+\Pi_{\alpha\beta}(s) \ . \label{Pback}
\end{equation}
\noindent

 The same constraints from unitarity and analyticity still apply as
given by eqs. (\ref{gs},\ref{impi},\ref{repi}), i.e., \Pii\ is given by
the same formulas as before although the interpretation of the coupling
constants now refers to the background.

We can thus add a  background term to each resonance by doubling the dimension
for the propagator, such that for the case of one resonance and a background
one writes a   $2\times 2$ inverse propagator matrix

\begin{equation}
 P^{-1}_{\alpha\beta }(s)=
 \left(\begin{array}{cc} m^2_0-s & 0  \\
0 & \pm M^2 \end{array} \right)    +\Pi_{\alpha\beta}(s) \ . \label{Presback}
\end{equation}
\noindent

The same formulas for \Pii\ still apply, but since \Pii\  has
off diagonal terms the background mixes with the resonance in a complex
although specified way just as in the two resonance case.
An important special case appears when the background
couplings are proportional to the resonance couplings, as is the case
when one assumes the same flavour structure for resonance and
background ($g_{\alpha i, backg}=\alpha g_{\alpha i,reson.}$).
In this case when one  algebraically reduces the  dimension of the propagator
matrix one gets the same resonance formula, but with
a modified $G_{\alpha i}(s)$.
With the same original $F_i(s)$ for resonance and background one needs only to
substitute
$g_{\alpha,i}^2 \to g_{\alpha,i}^2 [1 \pm
 \alpha^2 (m^2_{0,\alpha}-s)/M^2] $. This new linear factor
implies that there is a zero in the amplitudes, which we identify with
the Adler zeroes near $s=0$. Thus by  adding  the resonance and contact term
with a definite relative weight one introduces
the Adler zeroes\cite{adler} needed from current algebra.
We let  the Adler zeroes be at $s=s_{A,i}$.  All this is accomplished by the
simple substitution:

\begin{eqnarray}
 g_{\alpha i}^2 F^2_i(s) &\to &
\gamma_{\alpha i}^2 (s-s_{A,i})F_i^2(s), \ \ {\rm or}\\
 G_{\alpha i}(s)G_{\beta i}(s) &\to &
\gamma_{\alpha i}\gamma_{\beta i} (s-s_{A,i}) F^2_i(s)
\frac {k_i(s)}{\sqrt s} \theta (s-s_{th,i}), \label{gg}
\end{eqnarray}

\noindent where we now have introduced dimensionless coupling constants
$\gamma_{\alpha i}$, which are related by flavour symmetry eq.(\ref{flavour2}).

One can of course in a similar way include  more complicated dynamics
coming from $t$- and $u$-channel exchanges. By making a partial wave projection
of the Born term one first determines the $G_{\alpha i}(s)$ and
the function, which should replace $M^2$
or $m_0^2-s$ above, which will contain logarithms etc. Then
one sums the iterated higher order diagrams by the above
unitary formalism.

\vskip 0.6cm

{\bf 2.6 The UQM when two pseudoscalars are produced by some other reaction.}
\vskip 0.4cm

Much of the physics of $\pi\pi$ and $K\pi$ systems can be learnt from other
reactions than the two-body reactions discussed above. Such reactions are
$\Upsilon'\to\Upsilon\pi\pi$, $K\to \pi\pi e\nu$ or central production of
$\pi\pi$ in $pp$ collisions.
 The UQM can easily be modified in order to be applicable
also to such processes. All one needs to do
(provided the strong interactions in the final state is only
between the two pseudoscalars)
is to replace  the vertex functions $G_{\beta i}(s)$
at one of the two vertices by some other
real functions $C_{\beta p}(s)$, parametrized in an appropriate way for the
production, while
the propagator matrix including the bare masses
and the $\Pi_{\alpha\beta} (s)$ functions, and
$G_{\beta i}(s)$ at the second vertex for the decay remain as above:

\begin{equation}
 {\cal F}_{ip}(s) =   \sum_{\alpha\beta}
G^\dagger(s)_{i\alpha}
P_{\alpha\beta}(s) C_{\beta p}(s) \ , \label{aprimeij}
\end {equation}

This guarantees that the Watson final state theorem, and unitarity in general,
is automatically satisfied and the Adler zeroes will appear
only at one vertex in the remaining $G_{\alpha i}(s)$.

I believe this formalism should be much easier to apply,
and physically more transparent, than e.g., the formalism
of refs.\cite{bugg,morgan} where one multiplies the whole PWA's by real
functions $\alpha_i^p(s)$. I have made some initial
calculations along these lines, and seen
that it is easy with appropriate choices of the
$C_{\beta p}(s)$ to obtain the \fs\ either as a dip or as a peak in the
cross section. But more quantitative applications are left for further work.

\vskip 0.6cm
\eject

{\bf 2.7 Comparison with the N/D method and the K-matrix.
Physically acceptable analyticity.}
\vskip 0.4cm

For a single resonance and one threshold our function $G^2(s)$ is the  N
function, while our $P^{-1}(s)$ is the D function of the N/D method. For the
more general case  of many channels and many resonances
there is no clear connection between the parameters
in eq. (\ref{aij}) and those of a matrix form of N/D\cite{bj}.
But, much of the same
philosophy of N/D methods is also present in the UQM:
Given a model for $G_{\alpha i}(s)$
analyticity and unitarity constrain  the form of the propagator
$P_{\alpha\beta}$.

As to the relation between the K-matrix formalism and the UQM, one  obtains
the K-matrix corresponding to the matrix (\ref{aij}) by simply putting
$Im\Pi_{\alpha\beta}(s)=0$:

 \begin{equation}
  \hat K_{ij}(s) =  (\rho_i\rho_j)^{1/2}K_{ij}(s) =
\sum_{\alpha\beta} G^\dagger(s)_{i\alpha }
[m_{0,\alpha}^2 +Re\Pi_{\alpha\beta}(s) -s]^{-1}
G_{\beta j}(s) \ .  \label{kij}
\end {equation}

\noindent where $\rho_i =k_i /\sqrt s$.

The expression (\ref{aij}) is regained from the familiar formula:

\begin{equation}
  A = \hat K(1-i\hat  K)^{-1}  \ .  \label{akij}
\end {equation}

Note that the mass shift term $Re\Pi_{\alpha\beta}(s)$ remains in our
K-matrix. In particular, thresholds which  have not yet opened contribute
to $Re\Pi_{\alpha\beta}(s)$.
Thus the UQM can be looked upon as a particular way of parametrizing
the K-matrix in a way which is consistent with analyticity and dispersion
relations, whereby one has a simple
physical interpretation of the parameters.

 By adding the  $Im \Pi_{\alpha\beta}(s)$ terms to the propagator one
unitarizes
the model and sums over the imaginary parts of an infinite series of loop
diagrams (Fig.~1c).
But to add  the $Re\Pi_{\alpha\beta} (s)$ terms is equally important.
In many models such
terms from nearby closed thresholds are omitted, wheras I find it essential
to include a complete
set of nearby flavour related thresholds. Through these one includes the
mixing and mass shifts of the bare states with the continuum of meson pairs.

By omitting also the $Re\Pi_{\alpha\beta}(s)$ terms from the denominator
one obtains the bare matrix:

 \begin{equation}
   B_{ij}(s) = \sum_{\alpha} G^\dagger(s)_{i\alpha}
[m_{0,\alpha}^2  -s]^{-1}
G_{\alpha j}(s)\ .  \label{bij}
\end {equation}

Since the bare masses and the functions $G_{\alpha i}(s)$ determine
the PWA $A_{ij}$ through the scheme described above, there is a one to one
correspondence between the bare and the physical masses, once the functions
$G_{\alpha i}(s)$ are given.

One can gain some physical intuition for the UQM formalism, by viewing the
dispersion relation for $Re\Pi (s)$ as the limiting case
of the familiar second order perturbation formula for a mass shift
$\Delta E = \sum_i |<\alpha | H | i>|^2/(E-E_i)$. Each piece
of the continuum shifts the \qq\ state at the same time as
the continuum is mixed into the state,
with an amplitude $<\alpha | H|i>/(E-E_i)$.
The sum of the squared mixing amplitudes
$\sum_i |<\alpha | H| i>|^2$$/(E-E_i)^2$ again corresponds to the above
$-\frac d{ds}Re \Pi (s) $ $=-\frac 1 {\pi} \int Im \Pi (s')/[s-s']^2 ds'$.
 But, in contrast to second order
perturbation theory the dispersion relation formulas  are
 "exact" in the sense
that they solve the coupled channel model exactly.

It is important to emphasize that the function
$\Pi_{\alpha\beta}(s)$ must in general satisfy
a dispersion relation like the one in
eq.(\ref{repi}), whereby one automatically has
physically correct analytic properties for an
arbitrary form factor. But in the
literature one often finds violations of this rule.
A few examples should clarify this point:

If one would put $\Pi (s)\propto k(s)/\sqrt s$, i.e. make it proportional to
 the relativistic
phase space factor, (which would be reasonable for
Im $\Pi(s)$ if one had pointlike hadrons)
and use it both for the real and imaginary parts, one
would have a spurious pole and cut at $s=0$ in the physical region. The correct
procedure
is to calculate the real part from a dispersion relation, whereby one
gets the Chew-Mandelstam function, which is more complicated and
has a  logaritmic large $s$ behaviour (one needs one subtraction
constant because of the logaritmic
divergence). Then one has no spurious pole nor cut at $s$=0.

An even simpler example is given by the function $(4m^2-s)^{1/2}/s$, which
also has a spurious pole. But,
defining the function through the dispersion relation and its cut
(now one needs no subtraction constant) one gets $(4m^2-s)^{1/2}/s -2m/s$
i.e., one automatically subtracts the spurious pole at $s=0$
from the physical sheet. [On the
second sheet the pole remains, but does no harm:
There the function is $-(4m^2-s)^{1/2}/s -2m/s$.]
As a third example we take a finite cut as given by the function
$[(4m^2-s)(s-\Lambda^2)]^{1/2}$. The dispersion relation automatically
subtracts a polynomial, which guarantees that the function vanishes
asymptotically in the physical region:
One gets $[(4m^2-s)(s-\Lambda^2)]^{1/2}-s+2m^2+\Lambda^2/2$.

These added terms are not small nor insignificant.
They alter the model predictions
in a very crucial way, giving e.g., the sharp rise in
$Re\Pi (s)$ above the threshold (see Fig.~2).
 But as already mentioned,
in the literature one often encounters models, which either disregard the
real part entirely, or uses some variant of the
physically unacceptable forms discussed above.
Such forms might work within a limited energy region, if the
spurious singularities are very far away, but would certainly fail if one
considers several thresholds and a large energy region from threshold to
1.5 GeV, like in the present model.

\vskip 0.5cm

{\bf 3. Comparison with data on the lightest scalar mesons.}

\vskip 0.4cm

{\bf 3.1 Parameters and form factor}

\vskip 0.4cm

  As discussed in the previous section the PWA's of eq.~(\ref{aij}) are defined
once one has a model for the vertex functions $G_{\alpha i}(s)$
in eq.~(\ref{gs}) or (\ref{gg}) and the bare masses of eq.~(\ref{ps}).
For the latter it is natural to assume an "ideal" and isospin symmetric
structure such that the bare $u\bar u$ or $d\bar d$ mass $m_0$ is a free
parameter, while  the $u\bar s$ and $d\bar s$ bare masses are given   by
$m_0+m_s$ and the $s\bar s$ bare mass by $m_0+2m_s$, where $m_s$ is the
bare strange constituent mass.

For the bare couplings $\gamma_{\alpha i}$ we use the OZI rule giving
connected  flavour
symmetric couplings of eq. (\ref{flavour2}). The actual
numbers are given in Table 1 for general values of the pseudoscalar
mixing angle $\delta_P$, which measure the deviation from the ideal states. The
value of $\delta_P$ is fixed to $-54^\circ $ (which is equivalent to a
mixing angle $\theta_P=\delta_P+\cot^{-1} \sqrt 2 = -19^\circ $  for the
angle measuring the deviation from pure $SU3_f$ states). Such a value is
close to what has been measured in other contexts. (E.g. linear mass
matrix formulas give $\delta_P=-58^\circ $, quadratic $-46^\circ $, a recent
crystal ball experiment\cite{Xballeta} quotes $-52.2^\circ $,
Akers et al.\cite{akers} has $\delta_P=-53,7^\circ$, while  Gilman and
Kaufman\cite{gilman}, Baghi et al.\cite{baghi} and Donoghue et al. find
\cite{donoghue}
 $\delta_P\approx -55^\circ$. The results are not too
sensitive to its actual value as long as it is in this ball park.
All the above determinations assume the
mixing angle to be the same at the $\eta$
as at the $\eta '$ mass. This need not exactly be the case, since
mixing angles are in general mass dependent; cf. our
$\delta_S(s)$ below. This same simplifying assumption for $\delta_P$
is made also in this paper. With our
sign conventions the $\eta$ and $\eta '$ states are given by:

\begin{eqnarray}
|\eta\  > &=& -\sin \delta_P | u\bar u+d\bar d>/\sqrt 2 -\cos
\delta_P |s\bar s > \ , \cr
|\eta '> &=&  +\cos \delta_P | u\bar u+d\bar d>/\sqrt 2 -\sin
\delta_P |s\bar s > \ .
\end{eqnarray}

$$ \vbox {\halign { \hfil # \hfil &&  \hfil # \hfil \cr
\hline\cr\cr
& $\pi\eta$ & $K\bar K$&$\pi \eta'$ \cr
 $a_0$       & $-\sqrt 2 \sin \delta_P$& 1 & $\sqrt 2 \cos \delta_P$ \cr\cr
\hline \cr\cr
& $K\pi$ & $K\eta $&$K \eta'$ \cr
$K^*_0$ & $\sqrt \frac 3 2$&$-\sqrt\frac 3 2\cos (\delta_I-\delta_P)$& $\sqrt
\frac 3 2 \sin (\delta_I-\delta_P)$&\cr\cr
\hline \cr\cr
&  $\pi\pi$& $K\bar K$& $\eta\eta$&$ \eta\eta '$&$\eta '\eta '$\cr
$u\bar u +d\bar d$& $\sqrt 3$& 1 &$ \sin^2\delta_P$&
$ -\sin (2\delta_P)/\sqrt 2 $& $ \cos^2\delta_P$\cr
\ss\  & $ 0 $& $\sqrt 2$ &$ \sqrt 2\cos^2\delta_P$&
$ \sin (2\delta_P)$& $ \sqrt 2\sin^2\delta_P$\cr\cr
\hline \cr
}}$$
\vskip 0.2cm
{\small Table 1. The relative Clebsch-Gordan coefficients for
the coupling constants as given by the trace of eq.
 (\ref{flavour2}). The pseudoscalar mixing angle relative to the ideal
frame is fixed at $\delta_P=-54^\circ$,
while $\delta_I =\cot^{-1}\sqrt 2=35.26^\circ$ is the ideal mixing angle.}
\vskip 0.5cm

 More generally,
one could allow for contributions from disconnected, OZI rule violating
diagrams  for the bare couplings, like
$\gamma_{\alpha i}'= \gamma 'Tr[{\cal M}_{A_i}]
Tr [{\cal M}_{B_i}{\cal M}_{C_\alpha}] $
or $\gamma_{\alpha i}'=\gamma ''Tr[{\cal M}_{A_i}]
Tr [{\cal M}_{B_i}] Tr [{\cal M}_{C_\alpha}] $,
 which would introduce more parameters, and whereby
the bare couplings involving the neutral members
($\eta, \eta', f_0, f_0'$) would be different.
Note however, that our physical couplings do not
obey the OZI rule, because of
the nonzero value of the pseudoscalar mixing angle $\delta_P$, and because
the loops included in our formalism generate a scalar meson mixing angle
$\delta_S(s)$. It is  possible that most of the OZI
violation found is of this nature, coming from mixing in the mass matrix;
therefore for simplicity, we assume as a starting point that
the disconnected diagrams in the bare couplings can be put = 0.
Of course, if e.g. a nearby gluonium state would be present,
this assumption would have to be relaxed.

Since the data start only a few 100 MeV above the thresholds,
 they are not sensitive
to the exact positions of the Adler zeroes $s_{A_i}$ near $s=0$, provided
these are small enough or of the order 0.1 GeV$^2$. Thus we put all
$s_{A_i}=0$, except in the case of
the K$\pi$ and the $\pi\pi$ thresholds. For $\pi\pi$
we put it equal to the current algebra value
$s_{A_{\pi\pi}}=m^2_{\pi}/2$ (although
$s_{A_{\pi\pi}}=0$ would work just as well).
For $K\pi$ we let the fit find the best value,
which turns out rather large and negative $s_{A_{K\pi}}=-0.42$ GeV$^2$.
Current algebra including corrections from chiral perturbation
theory\cite{meissner} would predict this to be closer to zero. By
modifying the flavour symmetry prediction for the K$\pi$ coupling,
such that it is 10-15\%
larger than predicted by the Clebsch-Gordan coefficient of $\sqrt 3/2$ of
table 1, one could fit the data with a smaller $s_{A_{K\pi}}$.
Thus this determination of  $s_{A_{K\pi}}$ should not be taken too seriously.

\begin{figure}
\vspace{ 9.5cm}
\fcaption{The $K^*$ branching ratios into $K\pi$
for the resonances on the $K^*$  trajectory
as a function of the squared resonance mass. The
branching ratios fall as the square of the function
in eq.~(\ref{cutoff}) with $k_0=0.63$ GeV$^2$.
A similar exponentially falling  behaviour is seen\cite{goar}
for the  branching ratios into $\pi\pi$ for the
resonances on the $f_J$  and  $\rho_J$ trajectories.
See text.}
\end{figure}

Finally for the form factor $F_i(s)$ one assumes the simple Gaussian form:
\begin{equation}
 F_i(s) =exp[-k_i^2(s)/(2k^2_0) ]  \ . \label{cutoff}
\end{equation}

This, no doubt, is the most drastic assumption of the model. It is used because
it works, and because forms of this kind are obtained
in the quark pair creation model (QPCM)
\cite{olivier}. There, it is given by the overlap of the three $q\bar q$
wave functions of the three hadrons at the vertex times a matrix
element for
the $^3P_0$ quark pair creation. With wave functions of Gaussian shape one
gets a similar Gaussian factor, multiplied by a polynomial in $s$. The
cutoff parameter $k_0$ is then related to the hadron size $R$ through
$k_0=\sqrt 6/R$, from which with $R\approx 0.7$ fm
one would estimate $k_0$ to be of the order 0.6-0.8 GeV/c.
In the fit one finds $k_0=0.56$ GeV/c. Including another cutoff
parameter at  the quark pair creation vertex the Orsay group\cite{alain}
can account for a smaller cutoff parameter $k_0$.

There is another crude phenomenological argument for a form factor of this
form:
If one plots the elastic branching ratios for resonances on the
leading trajectory (i.e the $K\pi$ branching ratios
 for the $K^*_J$, see Fig.~3, or
the $\pi\pi$ branching ratio for the $f_J$ or $\rho_J$ trajectory\cite{goar})
 one finds that they fall
exponentially with squared mass or with $k^2$ as the square
of eq.(\ref{cutoff}) with $k_0 \approx 0.63$ GeV/c.
Of course this is not the same thing, but assuming that
angular momentum barriers approximately cancel in the branching ratio, that the
total widths grow much slower, perhaps linearly with $s$, and
that each resonance has an exponential form factor similar to (\ref{cutoff})
in their partial
widths to exclusive twobody channels etc., one can argue that there is some
experimental support for
such a shape and that the $k_0$ parameter is about the same for all mesons.

No doubt, $F_i(s)$ is in reality much more complicated than this. It should
in principle include all the left hand cuts. But since the left hand
singularities  lie much farther away than the unitarity cuts,
which we have included in  detail, one can expect that
the form (\ref{cutoff}) need not be too bad an approximation.

 In table 2 below the parameters of the model are summarized:
\vskip 0.1cm

$$ \vbox {\halign {  & # \hfil & #\hfil &   # \hfil \cr
&The over all dimensionless coupling constant \ \ \   \ & $\gamma$ & 1.1395 \cr
&The $u\bar u$ and $d\bar d$ bare $0^{++}$ mass & $m_0$ & 1.420 GeV \cr
&The constituent bare strange quark mass & $m_s$ & 100 MeV \cr
&The cutoff parameter & $k_0$ & 0.56 GeV/c \cr
&The Adler zero for $K\pi $ &$s_{A_{K\pi}} $ & -0.46 GeV$^2$ \cr
&The parameter enhancing the  $\eta \eta '$ couplings \ \ & $\beta$ & 1.6 \cr
}}$$
\vskip 0.2cm
{\small Table 2. The parameters of the model.
In addition, the pseudoscalar mixing angle  $\delta_P$ is
fixed to a conventional value of $-54^\circ$},
and the Adler zero for $\pi\pi$ is given the current algebra value $m^2_\pi/2$,
while for the remaining channels the Adler zeroes  $s_{A_i}$ are put = 0.
Other constants used in the model include the pseudoscalar masses and the
Clebsch-Gordan coefficients of Table 1.
\begin{figure}[p]
\vspace {20.2cm}
\fcaption{
(a) The  $K\pi$ S-wave phase shift (solid curve)
 and absorption parameter  $\eta*100$ (dotted curve) as functions of the
$K\pi$ center of mass energy when the
 model is fitted to the LASS
data\cite{LASS} shown by the round dots.
 The error bars of the data are of the same order
as the dots in the figure. This fit fixes four of the six parameters of
table 2: $\gamma,\  m_0+m_s,\ k_0,$ and $s_{A_{K\pi}}$. \\
(b) As in (a) but for the absolute
value of the S-wave amplitude.}
\end{figure}
\begin{figure}
\vspace{9cm}
\fcaption{ The $K\pi$ Argand diagram for the same fit as in Fig. 4.
The numbers indicate the $K\pi$ center of mass energy in MeV.}
\end{figure}

\vskip 0.5cm

{\bf 3.2. The $K^*_0(1430)$ and the $K\pi$ S-wave.}

\vskip 0.4cm

It is natural to start the data comparison  with the $K\pi$ S-wave,
since this is  the simplest to understand, having only one resonance,
the $K^*_0(1430)$, and in addition the experimental data are rather good.
The comparison of the fit to the LASS data\cite{LASS} is shown in Figs.~4a,b.
The older data of Estabrooks et al.\cite{penny} are very similar but have
larger error bars. The error bars of the LASS data are of the same magnitude
as the dots in the figures. Fig.~4a shows the S-wave phase shift, while
Fig.~4b shows the absolute magnitude of the amplitude (in Argand units). The
corresponding Argand diagram is shown in Fig.~5. As can be seen the model
has no difficulty in fitting the data rather well. These data essentially
fix four  of the  parameters in Table 2:
 $\gamma ,\ m_0+m_s ,\ k_0 $ and $s_{A_{K\pi}}$. We say essentially, since
some of the parameters are strongly correlated, in particular $k_0$ and
$\gamma$, and in practice because of the time consuming numerical integration
in eq. (\ref{repi}) one must keep $k_0$ fixed.

The parameters of the $K^*_0(1430)$ also turn out to be rather close
to the conventional ones (cf. Table 3). The BW mass is at 1349 MeV
where the phase shift passes $90^\circ$, and the
BW width is 398 MeV with negligible correction
from eq. (\ref{gamBW}), since $Re\Pi (s)$ is almost flat at 1350 MeV.
The nearest pole is also where one normally
expects it on the third sheet at $[Re (s)]^{1/2}
=1440$ MeV, with an imaginary part $-Im(s)/m= 320$ MeV (Since the $K\eta$
coupling almost vanishes this threshold is very weak and, in fact, there is
an image pole on the second sheet at almost the same position).

\vskip 0.5cm

\begin{figure}[p]
\vspace {20cm}
\fcaption{(a) The $\pi\eta$ S-wave Argand diagram as predicted by the model
when four of the parameters of table 2 are fixed by the $K\pi$ data of
Fig. 4 and $m_s =100$ MeV. Note the
sharply falling amplitude after the \KK\ threshold due to the strong absorption
to this channel, which partly explains the narrow \an\ shape. The numbers are
the center of mass energies in MeV.\\
(b) As in (a), but for the phase shift
(solid curve) and absorption parameter (dotted curve)
 plotted as a function of $\pi\eta$ center of mass energy.
}\end{figure}

\begin{figure}
\vspace {11cm}
\fcaption{The \an\ resonance shape (solid curve)
as predicted (not fitted) by the same model and
parameters as in Figs. 4 and 6, and compared with the
data of Armstrong et al.\cite{arms} from central production of $\pi\pi\eta$
in 300 GeV $pp$ collisions. The background (dashed curve)
is a rough estimate of the
reflection of other contributions to the experimental distribution and is added
to the model prediction. Note
that in spite of the inherently large width as measured by
$-Im\Pi (s)/m \approx 400$ MeV or by the imaginary part of the
pole position the predicted peak
width is rather narrow $\approx 100$ MeV.}
\end{figure}

{\bf 3.3. The $a_0(980)$ and the $\pi\eta $ S-wave.}

\vskip 0.4cm

When turning to the $\pi\eta$ data for the $a_0(980)$ there is only one
new parameter, $m_s$, whose value one expects to be close to 100 MeV. And in
fact, the fit allows for this value, therefore it is fixed at 100 MeV. The
value
100 MeV is slightly lower than what is
usually quoted in the light sector (e.g.,
the mass spittings of $\phi -K^*$ or $K^* - (\rho ,\omega)$ is about
120 MeV.) But, part of this splitting is renormalized by the unitarity
effects we include, and in the fits to the other light multiplets\cite{UQM}
values close to 100 MeV were found.
 A better comparison is with the $D_s -D$
splitting of $99.1\pm 0.6$ MeV, since here unitarity effects are smaller.

The data for the $a_0 (980) $ is much poorer than for the $K^*_0(1430)$.
In Fig.~6a the Argand diagram for the $\pi\eta \to \pi\eta$ S-wave is shown,
and  Fig.~6b shows the
 phase shift with the absorption parameter.  Fig.~7 shows the the prediction
for  the $a_0(980)$ peak compared with the data of Armstrong et al.\cite{arms}
of centrally produced $\eta\pi\pi$ in $pp$ collisions.
It is significant that the model predicts a rather narrow peak at the
right mass. This low  mass comes out right since the mass shift is the largest
at the threshold, and the
three thesholds for the $a_0$: $\pi\eta ,\ K\bar K ,\ \pi\eta'$ (Fig.~2b) lie
close to each others, while
 the corresponding ones for $K^*_0$: $K\pi ,\ K\eta ,\ K\eta'$, (Fig.~2a) are
much more spread out in energy and the middle one, $K\eta$, almost
decouples.
Again the rather narrow peak structure of about 100 MeV seems at first very
surprising, considering the very large
imaginary part (cf. in Fig.~2b, Im$\Pi (s)/m \approx$ 400 MeV). This is
due to several contributing  effects, which we list in their order of
importance:

\medskip

i) The large negative slope of $Re\Pi (s)$ below the \KK\ threshold,
which according to eq.(\ref{BW}) renormalizes the BW width by a large
factor, which is close to 5 at $\sqrt s=.98$ GeV;

ii) The strong absorption above the \KK\ threshold
(cf. the Argand diagram of Fig.~6a) making the amplitude drop fast in
magnitude above this threshold;

iii) The Adler zero at $s=0$ reduces the
amplitude somewhat near the threshold.

\medskip
See also the discussion of pole position, Sec.~3.5, for another way of
understanding the narrow peak.

\begin{figure}[p]
\vspace {19.8cm}
\fcaption{(a) The Argand diagram of the $\pi\pi $ S-wave as predicted
(not fitted)
by the model when 4 of the 6 parameters of table 2 are fixed by the
$K\pi$ data of Fig. 4 and $m_s=100$ MeV (as in Figs.6-7) and $\beta=1.6$.
The numbers are the center of mass energies in MeV. \\
(b) The same prediction as in (a) for the
 $\pi\pi$ phase shift (solid curve) and absorption
parameter $\eta *100$ (dashed) as function of the center of mass energy,
 and compared with the CERN-Munich\cite{otts},
Cason et al\cite{cason}, Grayer et al.\cite{grayer} and some low energy
experiments\cite{rosselet}. }
\end{figure}
\vskip 1.5cm

{\bf 3.4. The $f_0(980)$ and $f_0(1300)$ and the $\pi\pi$ S-wave.}

\vskip 0.4cm
Finally we turn to the much more complicated $\pi\pi$ S-wave
shown in Fig.~8 together with the CERN-Munich\cite{otts},
Cason et al.\cite{cason}, Grayer et al.\cite{grayer} and some lower energy
experiments\cite{rosselet}. Here we
have two resonances with complex energy dependent mixing. Now essentially
all of our parameters except one are already
fixed by the $K\pi$ and $\pi\eta$ data.
But even without new parameters,
  the prediction for the phase shift is not too bad below 1.1 GeV.
The additional parameter $\beta$ is needed
in order to fit the phase shift above this energy,
which otherwise comes out about 40$^\circ$ too low. We chose this to be
a factor in front of
the $\eta\eta '$ couplings, which thus violate the flavour symmetry
predictions of Table 1, being increased  by 60\%. This
determines our final parameter $\beta=1.6$. Of course this does not
necessarily mean that this is the right choice. Equally well one could
think of other effects, left out of the model, which could be faked by
$\beta$. These include higher thresholds which enter at about the same energy
as
$\eta\eta '$ in the 1.5 GeV region such as vector-vector or
pseudoscalar-axial thresholds etc., or an  effect due to
the  higher mass scalar resonances seen in
this region.

\begin{figure}[p]
\vspace{21cm}
\fcaption{
 The real (a) and imaginary (b) parts of the functions
$m^2_{\alpha\beta}(s)=m^2_{0\alpha}\delta_{\alpha\beta}+\Pi_{\alpha\beta}
(s)$ for the \uu\ and the \ss\ resonances  before diagonalization of the
mass matrix. Note that the \ss\  running mass $Re\ m^2_{22}(s)$ dips
strongly towards the \uu\  running mass $Re\ m^2_{11}(s)$
near the \KK\ treshold.
}
\end{figure}

\begin{figure}[p]
\vspace {18.5cm}
\fcaption{
 The real (a) and imaginary (b) parts of the eigenvalues of the
mass matrix  $m^2_{\alpha '}(s)$ after diagonalization, and (c)
the scalar mixing angle
$\delta_S(s)$. The crossing points with $s$ (dashed) in  (a) give the BW
masses.
Note that $\delta_S(s)$ is almost real at the energies of the BW masses, and
has a large imaginary part in the \KK\ threshold region. For low
energies the mixing is nearly ideal ($\delta_S=0^\circ$), wheras above 1.1 GeV
one has nearly pure $SU3_f$ eigenstates ($\delta_S=-35.26^\circ$).
 At the lighter BW  mass $\sqrt s =0.86$ GeV
one has an extremely  broad (880 MeV) near \uu\
BW resonance (the "$\sigma$") while at
1186 MeV one has a near $SU3_f$ octet BW resonance.  On the other hand, when
one
analytically continues to the nearest pole, these
combine linearily to give one very "narrow" almost pure \ss\
state [the \fs ], which lies on the second
(not third) sheet. See text for discussion.
}\end{figure}

In Figs.~9 and 10 the running masses are shown before, respectively after
diagonalization of the mass matrix. Note that in the region of the two
resonances the real parts of the two diagonal running masses are nearly
degenerate (Fig.~9a) due to the mass shifts from the \KK\ threshold. Therefore
the mixing angle (Fig. 10c)
between the two resonances changes rapidly at the \KK\
threshold from being nearly
ideal $\delta_S\approx 0^\circ$ below 900MeV to
$\delta_S\approx -35.26^\circ$ where the
states are close to being  $SU3_f$ eigenstates.
Note that the large value of $-Im (m^2_1)$ results in  a very large BW
width (880 MeV)
of the first BW resonance ("the $\sigma$").
Fig.~10a,b shows the eigenvalues, with the BW masses given by the crossing
points of the real parts with $s$. In Fig.~10 the \fs\ does not appear as a BW
resonance, but it is hidden under the complex and rapid energy dependent
mixing.
It is another linear combination of the two states, which appears when one
analytically continues to the nearby poles.

\vskip 0.5cm

{\bf 3.5. Pole positions and BW resonance masses and  widths.}

\vskip 0.4cm
When one asks the question, where are the poles of the resonances in this model
with large imaginary parts, one is tempted to say: Who cares where the
poles end up in the complex multisheeted structure of the $s$-plane!
Each resonance may require many poles on different sheets for a full
description
of its pole structure.
 On the other hand one has a very simple bare spectrum, and a rather
simple description of the data.
The bare masses are strongly shifted nonperturbatively to some $s$
 values and sheet numbers,
whose  positions depend sensitively on the precise positions of where
the thresholds are. Do we really learn anything fundamental from knowing the
precise pole positions, apart from having numbers to put into the PDG tables?
I am inclined to answer: not very much, they are of secondary importance!
Of course, the experimental data,
resonance shapes and threshold positions are very important in order to find
the correct solution with a reasonable bare spectrum together with the other
model parameters.
But the actual positions of the poles depend more on the positions of the
thresholds and on Clebsch Gordan coefficients, than on the model  parameters.

The actual pole positions do tell us something of the analytic structure of
the solution found, and can throw some light on resonance  shapes.
E.g., the fact that the nearest \an\ and \fs\ poles are found on the second
sheet, although their Re $(s_{pole})$ are above the \KK\ threshold, throws some
light on why these are narrow. Normally with two open thresholds the
nearest resonance
pole is expected on the third sheet. The fact that it often turns up
 on the second sheet explains partly their shapes:
The nearest place to the pole in the physical region is at the threshold,
since one must go around the \KK\ cut to see the pole. Therefore, a narrow peak
appears  at the threshold
in spite of a very large imaginary part of the pole position, and in spite
of the fact that  Re$(s_{pole})$ is about 90 MeV above the \KK\ threshold.
 Of course, as explained above one can  also
understand the narrow \an\ by the arguments given earlier by the structure
of the amplitudes in  the physical region  at Re$(s)+i\epsilon$.

In Table 3 I list the  Breit-Wigner masses, widths (in MeV) and mixing angles
together with the nearest pole positions, the  Riemann sheet of the poles and
mixing angles at the poles. Note that the $a_0$ and $f_0$ lie on the
second sheet although the \KK\ threshold is open at the pole mass,
as was already discussed above.  Similarily
the second $f_0$ lies on the 3rd sheet although the $\eta\eta$ threshold is
open, and one normally would expect it at the 5th sheet.  The $a_0$
width in the $\Gamma_{BW}$ column is  the peak width since it is  not
possible to define a BW width.

The mixing angle $\delta_S$
is such that at  the BW mass of the first $f_0$, whose BW
width is enormous, 880 MeV, $\delta_S$ is small, i.e. one has an almost
pure \uu\ state with a little mixture of \ss . The sign of this mixing
such that it is mainly an $SU3_f$
singlet state. This BW resonance we  interpret as the "$\sigma $". On the
other hand, when one analytically continues to
the pole with lowest mass one has another linear combination
of the two resonances: A  narrow almost pure \ss\ state (but with a large
$i$Im$\delta_{S,pole}=i39^\circ $). This is our $f_0(980)$.
See also our discussion in Sec 2.3 of how two
broad resonances can make one narrow. The heavier $f_0(1300)$ is  close to an
SU3$_f$ octet both at the BW mass and at the pole position. Also the
mass of the heavier state is similar both at the pole (1202 MeV) and when
measured as a BW resonance (1186 MeV).

$$ \vbox {\halign {& #\hfil&\hfil#\hfil&\hfil#\hfil&\hfil#\hfil&\hfil#\hfil
 &\hfil#\hfil&\hfil#\hfil&\hfil#\hfil&\hfil#\hfil \cr
\cr
Resonance\ \ &  $m_{BW}$ &  $\Gamma_{BW}$  & $\delta_{S,BW}$ &
$(Re\  s_{pole})^{\frac 1 2}$
 & $\frac {-Ims_{pole}}{m_{pole}}$ & Sheet nr &$\delta_{S,pole} $\cr
$a_0$ & 987  & $\approx  100^{\dagger)}$ &               & 1084  & 270 &II \cr
$K^*_0$& 1349&     398        &                      & 1441 & 320 &III\cr
Lighter $f_0$& 860& 880         & $-9^\circ +i8.5^\circ$ & 1006 & 33.7&II &
$0.4^\circ +i39^\circ$ \cr
Heavier $f_0$& 1186& 350       & $-32^\circ +i1^\circ$ & 1202 & 338 &III &
$-36^\circ+2^\circ$ \cr
}}$$
\vskip 0.2cm
{\small Table 3. Breit-Wigner masses, widths (in MeV) and mixing angles
together with the nearest pole position parameters, Riemann sheet number and
mixing angles at the poles.
 Except for the $\delta_{S,BW}$ at the light $f_0$
the three other mixing angles are given with respect to the \ss\ state.
The heavy $f_0$ is thus close to an $SU3_f$ octet, for which
$\delta_S=-35.26^\circ$, while the lighter BW $f_0$  is a near
\uu\ state (the $\sigma$). But at the poles one has other linear combinations
of the two resonances such that  the lighter is a narrow almost pure
\ss\ state, the $f_0(980)$.  \\
$^{\dagger)}$ The $a_0$
width in the $\Gamma_{BW}$ column is the peak width as it is here not
possible to define a BW width. See text.}
 \bigskip

Our solution suggests a novel way to resolve the old question: Where is the
$\sigma$ meson? From the low energy side the $\pi\pi$ interaction looks
very much as if there  would only be  an extremely
broad (880 MeV) BW resonance with a mass of
860 MeV, and with a composition near \uu . The fact that the running mass is
not constant makes the effective $\sigma$ mass slightly vary depending on what
energy one is sensitive to. It is interesting that this  prediction of the
$\sigma\to\pi\pi$ width is close to the values
predicted in the linear $\sigma$
model and in extensions of the Nambu--Jona-Lasinio  model\cite{hatsuda}.

On the other hand as one approaches the \KK\ theshold the two resonances
experience a dramatic $s$-dependent  mixing, such that at the \KK\ threshold
the relevant linear combination appears as an almost pure \ss\ pole
[the $f_0(980)$]. We find the pole to be a little above the threshold and on
the second sheet, with its precise position rather sensitive to model
parameters. But, it
is always a few MeV from the \KK\ threshold.

\vskip 0.5cm

{\bf 4. Concluding remarks.}
\vskip 0.4cm

In 1972 Feynman\cite{lipkin} wrote in connection with the problem of how to
treat virtual hadronic loops, needed by self consistency in the quark model,
that  "it is the most important problem in the theory of strong interactions",
but continued that
"no calculation of such virtual strong interactions in any problem has ever
been successful". He thought this
 "dead end is a result of lack of imagination of how to get further".

No doubt, these were strong words, and certainly we have made a lot of progress
to understand virtual hadronic states since 1972, but the progress has been
rather slow, possibly because too many theorists really thought that one
had reached a "dead end", and looked for other approaches guided by QCD.
Thereby the nonperturbative methods and related problems
of the sixties were "swept under the rug".

Today it seems more and more obvious that attempts where one, starting
directly from the QCD Lagrangian, tries to solve the nonperturbative
light hadron spectrum, has come to another "dead end", in spite of the
heroic attempts of e.g., lattice QCD.
This seems all the more obvious, with the present solution of the scalar nonet
at hand, which I believe to be close to the true solution,
although certainly  improvements can and should be done. To disentangle all the
complicated threshold singularities,
shifting the poles to some strange corners of  the
the many-sheeted structure of the complex $s$ plane,
and then try to understand why
some peaks like the \an\ or \fs\ are narrow in spite of their large
couplings, seems to be  an impossible task to resolve directly
from the QCD Lagrangean. It would require
 inclusion of fermionic loops and too much
computing power,
if one would not know how to proceed through some intermediate steps.

Such an intermediate step which I believe should  be useful, is  to
use QCD to calculate, or at least understand qualitatively, the parameters
 used in the UQM, i.e.,
the overall coupling constant $\gamma$, the cutoff parameter  $k_0$ and
the bare masses, when the mass shifts and
couplings to the lightest thresholds have been unfolded.

A few comments are in order as to the meaning of the bare masses found in this
paper. The value of $m_0$ depends on  the cutoff parameter
$k_0$ and on the number of thresholds, which have been included.
The larger the cutoff and the more thresholds included in the UQM, the larger
will $m_0$ be. Furthermore, if one also includes tadpole graphs
with hadron loops,
these have no imaginary part, but would contribute to $Re\Pi$,
 and shift $m_0$ down, partly cancelling the mass shift from the
thresholds. Thus the bare mass $m_0$ is model dependent.  On the other hand,
the bare constituent strange quark mass $m_s$ is less
dependent on distant singularities,
since in the mass difference $m_0(s\bar s) -
m_0(u\bar u)=2m_s$  the above effects essentially cancel. Similarily
 OZI rule violating mixing between resonances ($\delta_S$)
is less  dependent on distant thresholds, since
when one sums over complete sets of F-coupled and D-coupled flavour
related thresholds their contribution to off diagonal elements
in the mass matrix
is small\cite{UQM,geiger}. Only the overall bare mass scale
$m_0$ can depend strongly on
distant singularities.

The model we have found for the scalar nonet is, in principle,
not very different
than what one would assume for any of the other \qq\ nonets with finite
decay widths. For these one must always have an input mass scale $m_0$ and
a strange quark mass $m_s$ and an overall coupling parameter which,
together with flavour symmetry  describe the finite widths. Also the
cutoff parameter must, in principle, exist by self consistency,
although one implicitly
assumes that things do not depend on it. Our solution has the
same four parameters plus only two additional phenomenological parameters,
($\beta$ and $s_{A_{K\pi}}$), which
may be just artifacts of the things left out
of the model.
 Apart from the $\beta$ parameter flavour symmetry is broken in our
scheme most importantly by the pseudoscalar mass differences
and by the small $m_s$.

An important feature of the model is that the \an\ and the \fs\ have
 very large components of \KK\ (and also some $\eta' \pi $ or
$\eta\eta$ etc.) in their wave functions. We estimate
the \KK\ component to be dominant in the \an\ near the peak  being about 4-5
times larger than the \qq\ component. This does of course not mean that the
\an\ or the \fs\ are \KK\ bound states; they owe their existence to the
\qq\ component. But, it is important to note  that this will make
these states look "as if they were \KK\ states".
E.g., their widths to $\gamma\gamma$
will be reduced by this same factor of 4-5, and thus a reduced $\gamma\gamma$
width does not prove that they are \KK\ molecules, as has beed argued by
Close and Barnes and collaborators\cite{close}. One should not compare the \KK\
molecule model with a too naive \qq\ model, without the large virtual \KK\
component in the wave function.

 Finally a natural question many readers certainly ask: Why is it that
the scalar mesons are so much more sensitive to nonperturbative unitarity and
analyticity effects compared to the other light nonets? And a related
question: Why has not the solution presented here not been found previously?

To answer
 the first question one observes two special features for the scalars:

(i) The over all squared coupling
coupling constant $\gamma^2$ is about 5-6 times larger for the scalars than
for the vector nonet. This can be seen already by comparing the
$K^*_0(1430)$ width\cite{PDG} of 287 MeV, which is 5.7 times larger than
the $K^*$ width of 50 MeV. A better comparison taking into
account phase space etc. is found by determining a corresponding dimensionless
$\gamma^2$ parameter for the vectors from the width
expression: $(m\Gamma)_{K^*} = 3/2 \gamma^2 (4k^2) k/\sqrt s$, which also
gives a factor of about 5 smaller $\gamma^2$. Theoretically this is roughly
what one also expects from spin counting or SU6$_W$  symmetry, which would
predict a factor of 3. Thus the  pseudoscalar thresholds are about
five times stronger for the scalar mesons than for the vector mesons and
consequently unitarity effects are much more important.

(ii) The second feature which makes the thresholds more important for the
scalars is that these are S-wave thresholds. Therefore one has strong
square root cusps in $\Pi (s)$ with a discontinuous first derivative.
For P-wave and higher angular momenta this is smoothened out by the
angular momentum factor $k^{2\ell}$. Thereby the scalars are much more
sensitive to the threshold positions, especially when the resonance
turns up near a threshold as is the case for the \an\ and for the \fs .

As to the second question, why this solution has not been found previously,
one notes that (disregarding our first short note\cite{UQMscal}), no one has
tried to fit simultaneously the whole nonet taking into account all the
light pseudoscalar thresholds, putting in physically acceptable analyticity
properties etc.

\bigskip

{\it Acknowledgements.} Useful discussions with the Orsay group, in particular
with A. Le Yaouanc,
L. Olivier, O. P\`ene and J.C. Raynal, and with the meson team of the Particle
Data Group\cite{PDG}, especially Matts Roos,
are greatfully acknowledged. M. Sainio helped me to find references on
Adler zeroes and scattering lengths. I  also thank J. Soffer for
an invitation to Centre Physique Theorique, in Marseille, where part
of this work was written up.

\eject
{\bf Figure captions}
\bigskip

Fig. 1
The Born term for (a) the bare \qq\ resonances, for (b) the
contact terms, and (c) the loops summed by the functions $\Pi_{\alpha\beta}(s)$
in the inverse propagator.
\bigskip

Fig. 2.
 (a) The real and the imaginary parts of the function
$m^2(s)=m^2_{0i}+\Pi (s)$  for the
$K^*_0(1430)$ resonance plotted as a function of $\sqrt s$.
Note the strong cusp at the $K\eta '$ threshold and that
the $K\eta$ threshold essentially decouples because of the small
coupling constant.
The dashed curve is $s$, which crosses the running mass
Re$[m^2(s)]$ at the BW mass.

(b) As in (a) but for the
\an\ resonance. Note that the three thresholds have similar coupling
strengths and that they lie much closer together than
in (a). Therefore the large mass shift for \an . The dashed curve is $s$,
which crosses Re$[m^2 (s)]$ at the BW mass.

\bigskip

Fig. 3.
The $K^*$ branching ratios into $K\pi$
for the resonances on the $K^*$  trajectory
as a function of the squared resonance mass. The
branching ratios fall as the square of the function
in eq.~(\ref{cutoff}) with $k_0=0.63$ GeV$^2$.
A similar exponentially falling  behaviour is seen\cite{goar}
for the  branching ratios into $\pi\pi$ for the
resonances on the $f_J$  and  $\rho_J$ trajectories.
See text.
\bigskip

Fig. 4.
(a) The  $K\pi$ S-wave phase shift (solid curve)
 and absorption parameter  $\eta*100$ (dotted curve) as functions of the
$K\pi$ center of mass energy when the
 model is fitted to the LASS
data\cite{LASS} shown by the round dots.
 The error bars of the data are of the same order
as the dots in the figure. This fit fixes four of the six parameters of
table 2: $\gamma,\  m_0+m_s,\ k_0,$ and $s_{A_{K\pi}}$. \\
(b) As in (a) but for the absolute
value of the S-wave amplitude.
\bigskip

Fig. 5.
The $K\pi$ Argand diagram for the same fit as in Fig. 4.
The numbers indicate the $K\pi$ center of mass energy in MeV.
\bigskip

Fig. 6.
(a) The $\pi\eta$ S-wave Argand diagram as predicted by the model
when four of the parameters of table 2 are fixed by the $K\pi$ data of
Fig. 4 and $m_s =100$ MeV. Note the
sharply falling amplitude after the \KK\ threshold due to the strong absorption
to this channel, which partly explains the narrow \an\ shape. The numbers are
the center of mass energies in MeV.\\
(b) As in (a), but for the phase shift
(solid curve) and absorption parameter (dotted curve)
 plotted as a function of $\pi\eta$ center of mass energy.
\bigskip

Fig. 7.
The \an\ resonance shape (solid curve)
as predicted (not fitted) by the same model and
parameters as in Figs. 4 and 6, and compared with the
data of Armstrong et al.\cite{arms} from central production of $\pi\pi\eta$
in 300 GeV $pp$ collisions. The background (dashed curve)
is a rough estimate of the
reflection of other contributions to the experimental distribution and is added
to the model prediction. Note
that in spite of the inherently large width as measured by
$-Im\Pi (s)/m \approx 400$ MeV or by the imaginary part of the
pole position the predicted peak
width is rather narrow $\approx 100$ MeV.
\bigskip

Fig. 8.
 The Argand diagram of the $\pi\pi $ S-wave as predicted
(not fitted)
by the model when 4 of the 6 parameters of table 2 are fixed by the
$K\pi$ data of Fig. 4 and $m_s=100$ MeV (as in Figs.6-7) and $\beta=1.6$.
The numbers are the center of mass energies in MeV. \\
(b) The same prediction as in (a) for the
 $\pi\pi$ phase shift (solid curve) and absorption
parameter $\eta *100$ (dashed) as function of the center of mass energy,
 and compared with the CERN-Munich\cite{otts},
Cason et al\cite{cason}, Grayer et al.\cite{grayer} and some low energy
experiments\cite{rosselet}.
\bigskip

Fig. 9.
 The real (a) and imaginary (b) parts of the functions
$m^2_{\alpha\beta}(s)=m^2_{0\alpha}\delta_{\alpha\beta}+\Pi_{\alpha\beta}
(s)$ for the \uu\ and the \ss\ resonances  before diagonalization of the
mass matrix. Note that the \ss\  running mass $Re\ m^2_{22}(s)$ dips
strongly towards the \uu\  running mass $Re\ m^2_{11}(s)$
near the \KK\ treshold.
\bigskip

Fig. 10.
 The real (a) and imaginary (b) parts of the eigenvalues of the
mass matrix  $m^2_{\alpha '}(s)$ after diagonalization, and (c)
the scalar mixing angle
$\delta_S(s)$. The crossing points with $s$ (dashed) in  (a) give the BW
masses.
Note that $\delta_S(s)$ is almost real at the energies of the BW masses, and
has a large imaginary part in the \KK\ threshold region. For low
energies the mixing is nearly ideal ($\delta_S=0^\circ$), wheras above 1.1 GeV
one has nearly pure $SU3_f$ eigenstates ($\delta_S=-35.26^\circ$).
 At the lighter BW  mass $\sqrt s =0.86$ GeV
one has an extremely  broad (880 MeV) near \uu\
BW resonance (the "$\sigma$") while at
1186 MeV one has a near $SU3_f$ octet BW resonance.  On the other hand, when
one
analytically continues to the nearest pole, these
combine linearily to give one very "narrow" almost pure \ss\
state [the \fs ], which lies on the second
(not third) sheet. See text for discussion.

\end{document}